\documentclass[prl,twocolumn,amsmath,longbibliography]{revtex4-1}
\usepackage{graphicx}
\usepackage{bm}
\newcommand{\ket}[1]{|#1\rangle}
\newcommand{\bracket}[1]{\langle #1 \rangle}

\newcommand{\moire}{moir\'e }

\usepackage{siunitx}
\usepackage[usenames,dvipsnames,svgnames,table]{xcolor}
\usepackage[colorlinks=true,linkcolor=RoyalBlue,citecolor=RoyalBlue]{hyperref}
\usepackage{xcolor}

\AtBeginDocument{%
    \newwrite\bibnotes
    \def\bibnotesext{Notes.bib}
    \immediate\openout\bibnotes=\jobname\bibnotesext
    \immediate\write\bibnotes{@CONTROL{REVTEX41Control}}
    \immediate\write\bibnotes{@CONTROL{%
    apsrev41Control,author="08",editor="1",pages="1",title="0",year="1"}}
     \if@filesw
     \immediate\write\@auxout{\string\citation{apsrev41Control}}%
    \fi
}%

\begin{document}

\title{Topological charge pumping in twisted bilayer graphene}

\author{Yinhan Zhang}
\affiliation{Department of Physics, Carnegie Mellon University,
  Pittsburgh, PA 15213, USA}

\author{Yang Gao}

\affiliation{Department of Physics, Carnegie Mellon University,
  Pittsburgh, PA 15213, USA} 

\author{Di Xiao}

\affiliation{Department of Physics, Carnegie Mellon University,
  Pittsburgh, PA 15213, USA}

\date{\today}

\begin{abstract}
We show that a sliding motion between the two layers of a \moire superlattice induces an electric current and realizes a two-dimensional version of the topological Thouless pump when the Fermi energy lies in one of the minigaps. Interestingly, a chiral charge pump, namely, a transverse current induced by the sliding motion, is possible in twisted homobilayers. This result is confirmed by a concrete calculation of the adiabatic current in twisted bilayer graphene. Our work reveals an interesting link between mechanical motion and electricity unique to \moire superlattices, and may find applications in nanogenerators and nanomotors.
\end{abstract}

\maketitle

The emergence of long-wavelength \moire superlattices in van der Waals heterostructures has generated widespread interest in condensed matter physics. These superlattices can strongly modify the low-energy spectrum of charge carriers, giving rise to a wide range of spectacular quantum phenomena such as Hofstadter butterfly~\cite{hunt2013,dean2013,ponomarenko2013}, superconductivity~\cite{cao2018a,yankowitz2019,chen2019}, Mott insulators~\cite{cao2018,chen2019a}, and \moire excitons~\cite{seyler2019,tran2019,jin2019,alexeev2019}.  A notable recent advance in experimental techniques is the demonstration of in situ control of the \moire superlattice in rotatable heterostructures, including twisted bilayer graphene~\cite{cory2018} and graphene/BN heterostructures~\cite{finney2019}. These results point to the exciting possibility of dynamical control of quantum states in van der Waals heterostructures, in which the \moire superlattice should be treated as a time-dependent potential.

In a van der Waals heterostructure, the \moire superlattice is formed by incommensurate stacking of atomic layers.  A complete description of the stacking requires both the twist angle ($\theta$) and the relative shift ($\bm s$).  In principle, a variation in either $\theta$ or $\bm s$ will make the \moire potential time-dependent.  However, while varying the twist angle can lead to a drastic change of the \moire potential and hence a strong modification of the electronic structure, a relative shift does not change the overall shape of the \moire superlattice and the static electronic structure is shift independent (for small twist angles)~\cite{lopes-dos-santos2007,bistritzer2011,mele2011,lopes-dos-santos2012,moon2013}.  Because of this, so far little attention has been paid to the shift-dependence of the \moire superlattice.

\begin{figure}
\centering
\includegraphics[width=\columnwidth]{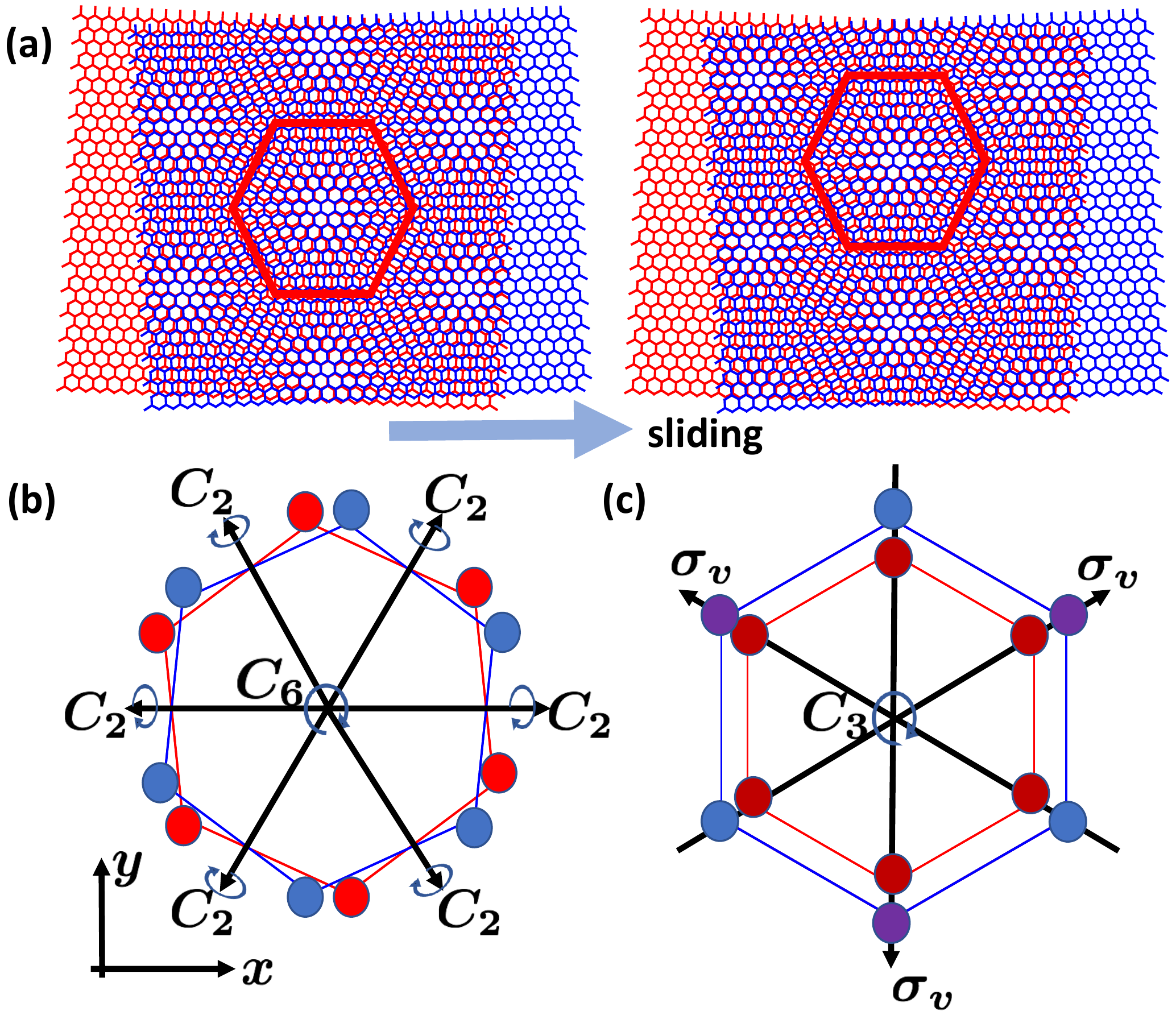}
\caption{(a) A relative shift in twisted bilayer graphene causes a transverse movement of the \moire superlattice. (b) Twisted bilayer graphene has $D_{6}$ symmetry. The red (blue) circle labels the carbon atom in the upper (bottom) layer. (c) Orientation-aligned graphene/BN heterostructure has $C_{3v}$ symmetry. The red, blue and purple circle label the carbon, boron and nitrogen atoms, respectively.}
\label{illustration}
\end{figure}

In this Letter we consider the dynamical consequence of the relative shift in a van der Waals heterostructure.  We show that a sliding motion between the two layers of a \moire superlattice will induce an electric current. In particular, when the Fermi energy lies in one of the minigaps, the system realizes a two-dimensional version of the topological Thouless pump~\cite{thouless1983}.  We first give a general argument of this phenomena by analyzing the motion of the \moire potential and the symmetry of the heterostructures.  Interestingly, we find that a chiral charge pump, namely, a transverse current induced by the sliding motion, is possible in twisted homobilayers [Fig.~\ref{illustration}(a)].  This result is further confirmed by a concrete calculation of the adiabatic current in twisted bilayer graphene, where pumping of Dirac electrons and the quantization of two-dimensional pumps are discussed.  Our result reveals an interesting link between mechanical motion and electricity unique to \moire superlattices, and may find applications in nanogenerators and nanomotors.

\textit{General argument}.---Since the concept of topological charge pumping is central to our analysis, we first give a brief account of its theory.  In 1983, Thouless showed that a cyclic variation of the Hamiltonian of a one-dimensional insulator can induce an adiabatic current, whose integral over a cycle period can be expressed as a topological number and thus must be quantized~\cite{thouless1983}.  This result can be intuitively understood in the simple case of a moving potential.  Consider a periodic potential $V(x, t) = V(x-v t)$ traveling at the velocity $v$.  Since the system is insulating, for sufficiently small $v$ inter-band transitions can be neglected, and we can track the motion of electrons by the movement of their Wannier centers.  As $V(x-vt)$ varies with time $t$, the Wannier centers must be moving along with the potential, producing a current $j = -nev$, where $n = N/a$ is the particle density with $N$ the number of filled bands and $a$ the lattice constant.  At $T = a/v$, the potential $V(x-vT)$ overlaps with $V(x)$, and the total number of pumped electrons is an integer $N$.

It turns out that the scenario of a moving potential can be readily realized in a \moire superlattice by a sliding motion between the two layers.  To see this, we model a periodic structure using a sinusoidal function $\sin(kx)$.  The superposition of two periodic structures, with one of them moving at the velocity $v$, can then be described by taking the average of two sinusoidal functions,
\begin{equation} \label{1Dmoire}
\begin{split}
&\frac{1}{2}[\sin(k(x- vt)) + \sin(k'x)] \\
&\quad = \sin \Bigl(\frac{k+k'}{2}x - \frac{1}{2}kv t\Bigr) \cos\Bigl(\frac{k-k'}{2}x - \frac{1}{2}kv t\Bigr) \;.
\end{split}
\end{equation}
If $k$ and $k'$ are similar to each other, then Eq.~\eqref{1Dmoire} describes a slowly varying envelope function (the second term) modulated by a fast varying function (the first term).  The \moire velocity $v_M$, which is the velocity of the envelope function, is given by $v_M = kv/(k-k')$.  

The above result can be generalized to two dimensions by switching to vector notation in Eq.~\eqref{1Dmoire}.  For small twist angle $\theta$ and lattice mismatch $\epsilon = (k - k')/k$, the \moire velocity is
\begin{equation} \label{moire_velocity}
\bm v_M = \frac{\epsilon\bm v + \theta\hat z \times \bm v}{\epsilon^2 + \theta^2} \;.
\end{equation}
If the Fermi energy lies in a gap opened by the \moire potential, we then expect that the sliding motion induces an adiabatic current
\begin{equation} \label{pump}
\bm j = -ne\bm v_M = -ne \frac{\epsilon\bm v + \theta\hat z \times \bm v}{\epsilon^2 + \theta^2} \;.
\end{equation}

It immediately follows from Eq.~\eqref{pump} that for twisted homobilayers ($\epsilon = 0$), a sliding motion will induce a current in the transverse direction [Fig.~\ref{illustration}(a)].  This can be understood by the following symmetry consideration.  In general, the sliding-induced charge pumping can be described by the phenomenological equation
\begin{equation} \label{symmetry}
j_\alpha = \chi_{\alpha\beta z}\tau_z v_\beta \;,
\end{equation}
where $\tau_{z}=\pm 1$ is the layer index which flips sign under mirror reflection in the $z$ direction.  The inclusion of $\tau_z$ is necessary because the sliding velocity, being the relative velocity between the two layers, has to flip sign under mirror reflection.  Hence, the charge pumping is controlled by a rank-$3$ tensor $\chi_{\alpha\beta z}$.

For a rank-3 tensor to be nonzero, inversion symmetry must be broken.  In addition, other point group symmetries will constrain the form of $\chi_{\alpha\beta z}$.  Let us consider two examples.  The first is a twisted homobilayer ($\epsilon = 0$), e.g., twisted bilayer graphene.  We begin with AA stacking and rotate one layer against the other with respect to the center of the honeycomb [Fig.~\ref{illustration}(b)].  It is clear that all three mirror symmetries are broken, i.e., the structure is chiral.  But it still has an out-of-plane $C_{6}$ axis and six in-plane $C_2$ axes.  The corresponding symmetry group is $D_{6}$, under which the only nonzero rank-3 tensor elements of the form $\chi_{\alpha\beta z}$ are $\chi_{xyz} = -\chi_{yxz}$.  A relative shift will obviously break the rotation symmetry.  However, for incommensurate structures, according to the equidistribution theorem of local geometries~\cite{massatt2017}, one should be able to find another place in the \moire superlattice where the centers of the two honeycombs overlap, and the same symmetry analysis applies.  The second example is an orientation-aligned heterobilayer ($\theta = 0$), e.g., graphene/BN.  We can see from Fig.~\ref{illustration}(c) that the symmetry group is $C_{3v}$, for which the relevant nonzero tensor elements are $\chi_{xxz} = \chi_{yyz}$.  In both examples, our symmetry analysis is in perfect agreement with Eq.~\eqref{pump}.

To estimate the size of the adiabatic current, we also need to know the charge density $n$.  Note that the low-energy effective Hamiltonians for common 2D materials, such as graphene and transition metal dichalcogenides, are of the Dirac type.  Counting the number of filled bands in this case may appear to be problematic since the spectrum of the Dirac Hamiltonian is unbounded.  As we will show later, the band filling should be counted from the charge neutral point.  Intuitively, this can be understood as the Dirac vacuum carries no charge nor current.  A similar situation has been discussed in the context of charge pumping in carbon nanotubes~\cite{talyanskii2001}. Therefore, for twisted bilayer graphene, if the first miniband is fully occupied, counting the spin and valley degeneracy, the particle density is $n = 4\times 2/(\sqrt{3}L_M^2)$, where $L_M = a/\theta$ is the \moire lattice constant.  The sliding-induced transverse adiabatic current is given by
\begin{equation} \label{j_tBLG}
j_\perp = nev/\theta = \theta \frac{8e}{\sqrt{3}a^2} v \;.
\end{equation}
Similarly, for graphene/BN, when the the first miniband is fully occupied the longitudinal adiabatic current reads
\begin{equation} \label{j_GBN}
j_\parallel = \epsilon \frac{8e}{\sqrt{3}a^2} v \;.
\end{equation}

\textit{Twisted bilayer graphene}.---To validate the above analysis, we now carry out a concrete calculation of the sliding-induced adiabatic current in twisted bilayer graphene.  Our starting point is the continuum model in Ref.~\cite{bistritzer2011}.  The real-space Hamiltonian for a single valley $\bm K$ is
\begin{equation} \label{Hamiltonian}
H_{\bm K}^{\theta}(\bm r) = -iv_F \bm\sigma \cdot \bm\nabla - \frac{k_\theta}{2} v_F \sigma^y \tau^z + [V(\bm{r}) \tau^+ + \text{h.c.}] \;,
\end{equation}
where the Pauli matrices $\sigma^\mu$ and $\tau^\mu$ operate in the sublattice and layer space, respectively.  The first two terms in Eq.~\eqref{Hamiltonian} describe the intralayer kinetic energy with $k_\theta = \theta k_D$, where
$k_D$ is the magnitude of the Brillouin-zone corner wave vector.  The last term is the ``potential'' energy arising from the periodic interlayer hopping, given by
\begin{equation}
V(\bm r) =\sum_{j=0}^{2}T_{j}e^{i\bm{Q}_{j}\cdot \bm r} \;,
\end{equation}
where
\begin{gather}
\bm Q_0 =0,\quad 
\bm Q_1= k _\theta (\frac{\sqrt{3}}{2},\frac{3}{2})\;,\quad 
\bm Q_2=k_\theta (-\frac{\sqrt{3}}{2},\frac{3}{2}) \;, \\
T_j = w\Bigl[\sigma^0+\cos(j\frac{2\pi}{3})\sigma^x+\sin(j\frac{2\pi}{3})\sigma^y\Bigr] \;,
\end{gather}
with $w=110$ meV being the nearest momentum hopping amplitude.  One can verify that $H_{\bm K}^{\theta}(\bm r)$ has $D_3$ symmetry.  The Hamiltonian for the other valley $\bm K^\prime=-\bm K$ can be obtained by a time reversal operation, i.e., $H^{\theta}_{-\bm K}(\bm r)=H^{\theta\ast}_{\bm{K}}(\bm r)$.  

\begin{figure}
\centering
\includegraphics[width=\columnwidth]{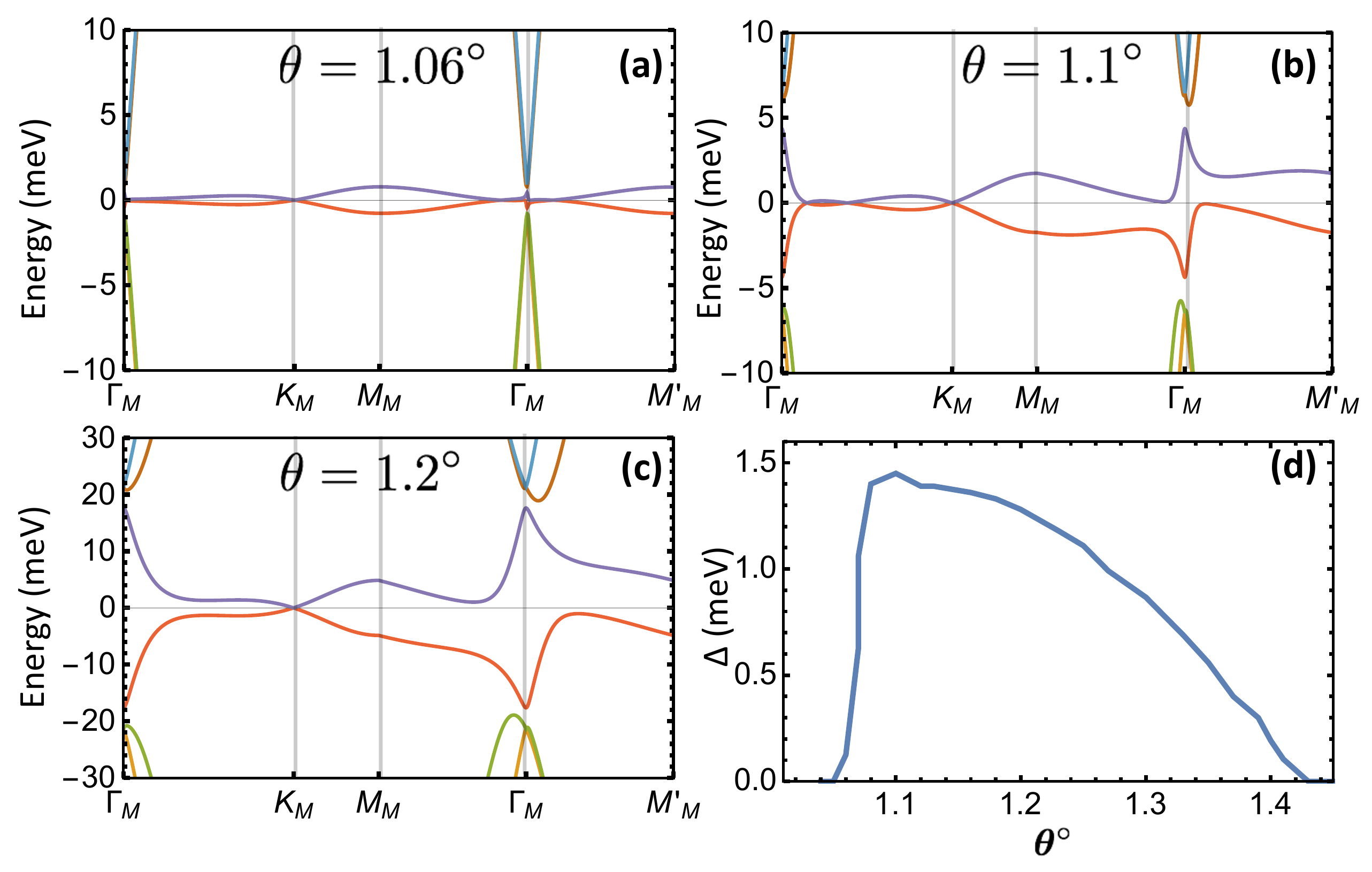}
\caption{(a)-(c).  The band structures of twisted bilayer graphene at (a) $\theta=1.06^{\circ}$, (b) $\theta=1.1^{\circ}$ and (c) $\theta=1.2^{\circ}$.
(d) The twist angle dependence of the global gap separating the first miniband to higher bands.}
\label{fig_dispersion}
\end{figure}

Next we introduce the shift dependence into the \moire Hamiltonian $H^{\theta}_{\bm K}(\bm r)$.  It has been shown in Ref.~\cite{bistritzer2011} that under a relative shift $\bm s$, the hopping matrices $\{T_1, T_2, T_3\}$ transform into $\{T_0, e^{-i\bm b_1 \cdot \bm s} T_1, e^{-i\bm b_2 \cdot \bm s} T_2\}$, where $\bm b_{1,2} = (2\pi/a)(1, \mp 1/\sqrt{3})$ are the reciprocal lattice vectors of a monolayer graphene.  Note that $\bm s\cdot \bm b_1=\bm Q_1 \cdot (\hat z\times \bm s/\theta)$ and $\bm s\cdot \bm b_2 = \bm Q_2 \cdot (\hat z\times \bm s/\theta)$.  Therefore, the \moire potential after a relative shift $\bm s$ can be written as 
\begin{equation} \label{potential}
V(\bm s; \bm r) = V(\bm r-\frac{1}{\theta}\hat{z}\times \bm s) \;,
\end{equation}
and the shift $\bm s$ enters into the \moire Hamiltonian $H^{\theta}_{\bm K}(\bm s;\bm r)$ 
via a gauge transformation
\begin{equation} \label{gauge}
H^{\theta}_{\bm K}(\bm s;\bm r) = e^{\frac{i}{\theta}\hat{z}\times \bm s \cdot \hat{\bm p}} 
H^{\theta}_{\bm K}(\bm r) e^{-\frac{i}{\theta}\hat{z}\times \bm s \cdot \hat{\bm p}} \;.
\end{equation}
Its band dispersion is independent of $\bm s$ and will be simply denoted by $\varepsilon_{n\bm k}$ in the following. 

A necessary condition to realize the Thouless pump is the existence of a global gap.  Figure~\ref{fig_dispersion}(a)-(c) show the band structure at three representative angles: $\theta=1.06^\circ$, $\theta=1.1^\circ$ and $\theta=1.2^\circ$.  We can see that there is a region of $\theta$ in which a global gap separating the first miniband from higher bands exists.  The global gap as a function of the twist angle is shown in Fig.~\ref{fig_dispersion}(d).  The gap reaches its maximum about 1.5 meV at the twist angle $\theta = 1.1^\circ$.  In the following, we shall assume that the Fermi energy lies in this gap, and set $\theta = 1.1^\circ$.  

\begin{figure}
\includegraphics[width=\columnwidth]{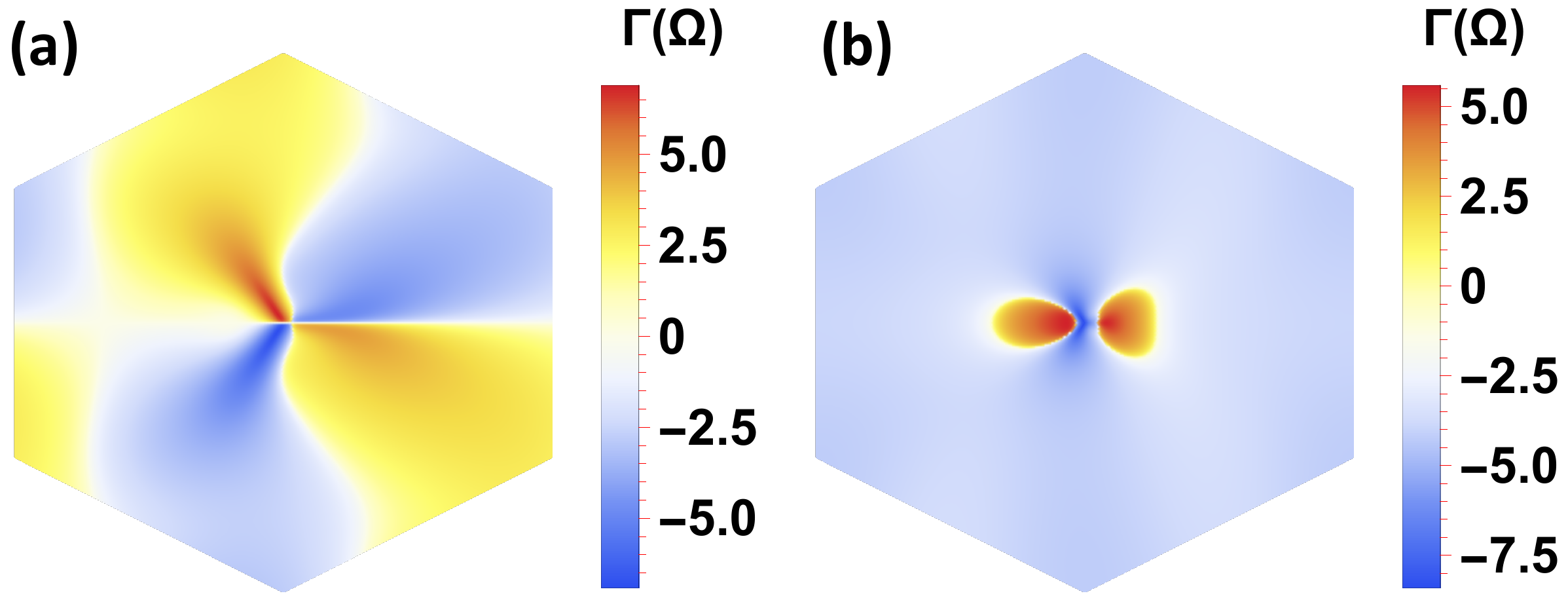}
\caption{The distribution of the total Berry curvature in log scale $\Gamma(\Omega_{k_is_j}) \equiv \text{sgn}(\Omega_{k_is_j})\ln(1+|\Omega_{k_is_j}|)$ at the twist angle $\theta=1.1^{\circ}$. (a) $\Gamma(\Omega_{k_xs_x})$; (b) $\Gamma(\Omega_{k_xs_y})$.}
\label{fig_BerryC}
\end{figure}

The pumping coefficient $\chi_{\alpha\beta z}$ is given by the summation of the Berry curvature of all occupied bands over the \moire Brillouin zone (MBZ)~\cite{thouless1983,xiao2010},
\begin{equation} \label{coefficient}
\chi_{\alpha\beta z}(\bm s) = -e\sum_{n \in \text{occ.}} \int_\text{MBZ} \frac{d^2k}{(2\pi)^2} \Omega^{n}_{k_{\alpha}s_{\beta}}(\bm k,\bm s) \;,
\end{equation}
where $\Omega^{n}_{k_{\alpha}s_{\beta}}(\bm k,\bm s)$ is the Berry curvature defined in the parameter space spanned by $\bm k$ and $\bm s$,  
\begin{equation}\label{kubo}
\Omega^{n}_{k_{\alpha}s_{\beta}}(\bm k,\bm s)=\sum_{m \neq n}\frac{v_{k_\alpha;nm}(\bm k,\bm s)v_{s_{\beta};mn}(\bm k,\bm s)+\rm{c.c.}}{\bigl(\varepsilon_{n\bm k}-\varepsilon_{m\bm k})^2}
\end{equation}  
with $v_{k_{\alpha};nm}(\bm k,\bm s) = \bracket{u_{n\bm k}(\bm s)|\partial_{k_{\alpha}}H^\theta_{\bm K}(\bm k,\bm s)|u_{m\bm k}(\bm s)}$ and 
$v_{s_{\beta};mn}(\bm k, \bm s)=\bracket{u_{m\bm k}(\bm s)|\partial_{s_{\beta}}H^\theta_{\bm K}(\bm k, \bm s)|u_{n\bm k}(\bm s)}$.  The wave function $\ket{u_{n\bm k}(\bm s)}$ is the eigenstate of the Bloch Hamiltonian $H^\theta_{\bm K}(\bm k,\bm s; \bm r)=e^{-i\bm k\cdot \bm r}H^\theta_{\bm K}(\bm s; \bm r)e^{i\bm k\cdot \bm r}$.  By making use of Eq.~\eqref{gauge}, one can show that both $v_{k_{\alpha};nm}(\bm k,\bm s)$ and $v_{s_{\beta};mn}(\bm k, \bm s)$ are independent of $\bm s$.  Consequently, we shall drop the $\bm s$-dependence of the Berry curvature $\Omega^{n}_{k_{\alpha}s_{\beta}}(\bm k,\bm s)$ and the pumping coefficient $\chi_{\alpha\beta z}(\bm s)$.  This can be also seen by noting that both the Berry curvature and the pumping coefficient are physical quantities, therefore they have to be invariant under the gauge transformation in Eq.~\eqref{gauge}.

The spectrum of the Hamiltonian $H_{\bm K}^{\theta}(\bm r)$ is unbounded.  Therefore the summation in Eq.~\eqref{coefficient} should extend to $n = -\infty$.  In practice, the dimension of the Hamiltonian matrix is determined by a truncation in the plane wave expansion of the eigenstates~\cite{bistritzer2011}.  We find that the Berry curvature quickly converges as more bands are included.  Figure~\ref{fig_BerryC} shows the distribution of the converged total Berry curvatures $\Omega_{k_{\alpha}s_{\beta}}(\bm k)=\sum^{+1}_{n=-\infty}\Omega^{n}_{k_{\alpha}s_{\beta}}(\bm k)$ where we have summed over all occupied bands up to the first miniband.   We can see that $\Omega_{k_{x}s_{x}}(\bm k)$ is odd under mirror reflection in the $y$ direction while $\Omega_{k_{x}s_{y}}(\bm k)$ is even.  After integrating over the Brillouin zone, only $\chi_{xyz}$ survives.  Taking into account of the spin and valley degeneracy, we find $\chi_{xyz} = -0.088 e/a^2$.  Compared with Eq.~\eqref{j_tBLG}, it corresponds to a charge density of approximately 4 electrons per \moire unit cell, confirming our earlier statement that the band filling should be counted from the charge neutral point.

What is the upper limit of the sliding velocity?  This is given by the adiabatic condition $\hbar \langle\partial_{t} H(t)\rangle\ll \Delta^2$, where $\Delta$ is the energy gap.  As the $t$ dependence of the Hamiltonian enters via the $\bm s$ dependence of the \moire potential, $\bracket{\partial_t H(t)}$ is approximately $w v /a$.  We thus have, for $\Delta\sim 1$ meV and $a = 0.246$ nm,
\begin{equation}
v \ll \Delta^2 a/(\hbar w) \approx 3~\text{m/s}.
\end{equation}
For $\theta = 1.1^\circ$ and $v = 10$ $\mu$m/s, which is well below the upper limit, we find the adiabatic current is 4 pA for a sample with a cross section of 1 $\mu$m.

Before closing, we briefly comment on the topological aspect of the 2D charge pumping process.  Our parameter space is spanned by two vectors, $\bm k$ and $\bm s$.  Since the \moire potential has the property $V(\bm s + \bm R; \bm r) = V(\bm s; \bm r)$, where $\bm R$ is a lattice vector of the graphene layer, the shift vector $\bm s$ lives on the torus spanned by the two primitive lattice vectors $\bm a_1$ and $\bm a_2$.  On the other hand, the momentum $\bm k$ lives on the torus spanned by the two primitive reciprocal lattice vectors $\bm G_1$ and $\bm G_2$ of the \moire superlattice.  Let us write $\bm s = x_1 \bm a_1 + x_2 \bm a_2$ and $\bm k = y_1\bm G_1 + y_2\bm G_2$.  Then,
\begin{equation}
\Omega_{k_\alpha s_\beta} = \frac{1}{(2\pi)^2} \sum_{ij} L^{\alpha}_{i} b^{\beta}_{j} \Omega_{y_ix_j} \;,
\end{equation}
where $\bm b_j$ and $\bm L_i$ are the reciprocal vectors of $\bm a_j$ and $\bm G_i$, respectively ($\bm a_i \cdot \bm b_j = \bm L_i \cdot \bm G_j = 2\pi \delta_{ij}$), and $\bm a_i$ and $\bm L_i$ are related by $\bm L_i = \hat z \times \bm a_i/\theta$.  The integral of $\chi_{\alpha\beta z}$ over a graphene unit cell $V_0$ then becomes
\begin{equation}
\begin{split}
&\int_{V_0} d^2s\, \chi_{\alpha\beta z} = -e \int_{V_0} d^2s \int_\text{MBZ} \frac{d^2k}{(2\pi)^2}\, \Omega_{k_\alpha s_\beta} \\
&\quad = -\frac{e\theta^2}{(2\pi)^2} \sum_{ij} L^\alpha_{i} b^\beta_{j} \int_0^1 d^2x \int_0^1 d^2y\, \Omega_{y_ix_j} \;,
\end{split}
\end{equation}
where we have dropped the summation over the band index for simplicity.  The integration of $\Omega_{x_iy_j}$ over $x_i$ and $y_j$ is quantized and is equal to $2\pi C_{ij}$, where $C_{ij}$ is the Chern number.  As long as the band gap does not close as $\bm k$ and $\bm s$ are varied, $C_{ij}$ should be a constant.  Recall that $\chi_{\alpha\beta z}$ is independent of $\bm s$, we finally have
\begin{equation}
\chi_{\alpha\beta z} = -\frac{e\theta^2}{(2\pi) V_0} \sum_{ij} L^i_\alpha b^j_\beta C_{ij} \;.
\end{equation}
Thus the 2D topological pump is characterized by four Chern numbers.

In summary, we have shown that the sliding motion in a \moire superlattice can realize a topological Thouless pump. A general expression for the adiabatic current has been derived.  Even though we have considered a rigid twist structure, our result applies even when there is lattice relaxation as can be seen from our topological consideration.  Actually, lattice relaxation in twisted bilayer graphene has been found to significantly enhance the minigap around the first magic angle~\cite{cao2018,carr2019}, which can push the upper limit of the sliding velocity even higher.  In addition, interaction induced gaps have been found at non-integer fillings such as 1/2, and topological charge pumping inside these gaps will be an interesting topic to explore~\cite{niu1984}.  Finally, we would like to emphasize that even though our discussion has been focused on adiabatic transport, the phenomenological equation~\eqref{symmetry} is quite general and should be valid even in metallic systems.  The chiral response of twisted homobilayers to a mechanical motion could be useful in nanogenerators and nanomotors.

\begin{acknowledgments}
D.X.\ thanks Ella Xiao for bringing his attention to the phenomenon of \moire speedup.  We are grateful to David Cobden, Ben Hunt, Eugene Mele, H\'ector Ochoa, Michael Widom and Matthew Yankowitz for fruitful discussions.  This work was supported by the Department of Energy, Basic Energy Sciences, Materials Sciences and Engineering Division, Pro-QM EFRC (DE-SC0019443).  Y.G.\ acknowledges the support from the Department of Energy, Basic Energy Sciences, Materials Sciences and Engineering Division (DE-SC0012509). D.X.\ also acknowledges the support of a Simons Foundation Fellowship in Theoretical Physics.
\end{acknowledgments}

\textit{Note added.}---Recently, we became aware of Ref.~\cite{fujimoto2019}, which obtained the same result of topological charge pumping in twisted bilayer graphene mainly using the tight-binding approach, while our analysis is base on the continuum model.

%

\end{document}